%% file: quperman.tex
\documentclass[sigconf]{acmart}

\usepackage{multirow}
\usepackage{xcolor}
\usepackage{tabularray}
\usepackage{hyperref}
\usepackage{subcaption}

\usepackage{tabularx}

\usepackage{framed}
\usepackage[most]{tcolorbox}
\newtcolorbox{bluebox}[1][]{
  colback=blue!5!white,
  colframe=blue!75!black,
  boxsep=2pt,      
  top=2pt,         
  bottom=2pt,      
  left=5pt,        
  right=5pt,       
  width=\linewidth 
}
\newtcolorbox{boxK}{
    colframe = cs-2!80!white,
    sharpish corners, 
    boxrule = 0pt,
    boxsep = 0pt,
    toptitle = 2pt,
    bottomtitle = 1pt,
    top = 4pt,
    bottom = 5pt,
    left = 5pt,
    right = 5pt,
    enhanced,
    fuzzy shadow = {0pt}{-1pt}{-0.5pt}{0.5pt}{black!20}, 
    fonttitle=\small\bfseries,
    fontupper=\small,
    title=Main Findings
}
\usepackage{blindtext}
\usepackage{balance}
\balance

\usepackage{svg}

\definecolor{main}{HTML}{5989cf}    
\definecolor{sub}{HTML}{cde4ff}     

\definecolor{firebrick}{HTML}{B22222}
\definecolor{yellowgreen}{HTML}{9ACD32}
\definecolor{royalblue}{HTML}{4169E1}

\begin{document}
\title[QUPER-MAn]{QUPER-MAn: Benchmark-Guided Target Setting for Maintainability Requirements}

\author{Markus Borg}
\orcid{XXX}
\affiliation{%
  \institution{CodeScene and Lund University}
  \city{Malmö}
  \country{Sweden}
}
\email{markus.borg@codescene.com}

\author{Martin Larsson}
\orcid{XXX}
\affiliation{%
  \institution{Lund University}
  \city{Lund}
  \country{Sweden}
}
\email{martin@mandella.se}

\author{Philip Breid}
\orcid{XXX}
\affiliation{%
  \institution{Lund University}
  \city{Lund}
  \country{Sweden}
}
\email{breidphilip@gmail.com}

\author{Nadim Hagatulah}
\orcid{XXX}
\affiliation{%
  \institution{Lund University}
  \city{Lund}
  \country{Sweden}
}
\email{nadim.hagatulah@cs.lth.se}

\renewcommand{\shortauthors}{Borg et al.}

\begin{abstract}
Maintainable source code is essential for sustainable development in any software organization. Unfortunately, many studies show that maintainability often receives less attention than its importance warrants. We argue that requirements engineering can address this gap the problem by fostering discussions and setting appropriate targets in a responsible manner. In this preliminary work, we conducted an exploratory study of industry practices related to requirements engineering for maintainability. Our findings confirm previous studies: maintainability remains a second-class quality concern. Explicit requirements often make sweeping references to coding conventions. Tools providing maintainability proxies are common but typically only used in implicit requirements related to engineering practices. To address this, we propose QUPER-MAn, a maintainability adaption of the QUPER model, which was originally developed to help organizations set targets for performance requirements. Developed using a design science approach, QUPER-MAn, integrates maintainability benchmarks and supports target setting. We posit that it can shift maintainability from an overlooked development consequence to an actively managed goal driven by informed and responsible engineering decisions.
\end{abstract}




\maketitle

\input{body.tex}

\balance
\bibliographystyle{ACM-Reference-Format}
\bibliography{references}

\appendix
\section{Questionnaire} \label{app:quest}
This appendix lists the questions in the body of the questionnaire, i.e., after providing background information about the research and the data management, and asking the respondent for informed consent. A complete description, including rich explanations, is available in the MSc thesis~\cite{breid_improving_2025}.
\begin{enumerate}
    \item[D1] In which domain(s) do you currently work?

    [a) Computer (Software), b) Telecommunication, c) Engineering/Architecture, d) Automotive, e) End-user perspective, f) Mobile Applications, g) Consulting, h) Internet/eCommerce, i) Finance/Banking/Insurance, j) Computer (Hardware), k) Education, l) Healthcare/Medical, m) Aerospace/Aviation, n) Government/Military, o) Business/Professional Services, p) Accounting, q) Logistics/Shipping, r) Entertainment/Recreation, s) Manufacturing, t) Corporate Communication, u) Biotechnology, v) Advertising, w) Food Service, x) Other]

    \item[D2] How many years of working experience in software development do you have?

    [a) 0-4 years, b) 5-9 years, c) 10-19 years, d) 20-24 years, e) 25 years]

    \item[D3] Which of the following best describes your current role?
    
    [a) Product development, b) System view/architect, c) Product planning, d) Quality assurance/system testing, e) End-user perspective, f) Operational Management, g) External business perspective, h) Maintenance evolution perspective, i) Legal perspective, j) Other]
    
    \item[D4] In a short sentence, describe your experience working with software requirements.

    [Free-text]
    
    \item[D5] How many software developers work in your company?
    
    [a) 1-4, b) 5-19, c) 20-49, d) 50-99, e) 100-199, f) 200-499, g) 500+]
    \item[L1] Please indicate your level of agreement with the following statements
    
    [Strongly disagree (1) -- Strongly agree (5)]
    \begin{enumerate}
        \item The software development in our organization is agile rather than plan-driven.
        \item Developers in our organization are supported by modern development tools.
        \item Systems for which our organization is responsible for maintaining the source code often carry high levels of technical debt.
        \item The development projects in the organization often have flexible deadlines.
    \end{enumerate}
    \item[L2] Please indicate your level of agreement with the following statements. [Strongly disagree (1) -- Strongly agree (5)]

    In our organization, it is important that\ldots
    \begin{enumerate}
        \item \ldots a system or computer program is composed of discrete components such that a change to one component has minimal impact on other components. (modularity)
        \item \ldots code assets can easily be used in more than one system, or in building other assets. (reusability)
        \item \ldots test criteria can be established for a system, product or component and that tests can be performed to determine whether those criteria have been met. (testability)
        \item \ldots it is possible to assess the impact on a product or system of an intended change to one or more of its parts, or to diagnose a product for deficiencies or causes of failures, or to identify parts to be modified. (analysability)
        \item \ldots a product or system can be effectively and efficiently modified without introducing defects or degrading existing product quality. (modifiability)
    \end{enumerate}
    \item[C1] Which of the following sub-dimensions are the most challenging for your organization? Select up to three options. 
    
    [a) modularity, b) reusability, c) testability, d) analysability, e) modifiability]
    \item[O1] Does your organization explicitly specify maintainability requirements in any format? If yes: Please describe how these maintainability requirements are specified. If no: Please explain the reasons for not specifying maintainability requirements. 
    
    [Free-text answer]

\end{enumerate}

\end{document}

%% file: body.tex
\section{Introduction} \label{sec:intro}
Software maintainability is a core responsibility in modern software engineering. As described in seminal work by Lehman and colleagues~\cite{lehman_laws_1996}, software must continually change to remain relevant and useful -- known as the first law of software evolution. Unfortunately, their second law states that the complexity of evolving software increases unless work is done to maintain or reduce it. This is an ever-present challenge, as organizations invest more into maintaining current systems than developing new ones and between 50\% to 80\% of project costs are directed at maintenance work~\cite{ernst_technical_2021}.

Maintainability concerns are commonly discussed under the Technical Debt (TD) umbrella. TD refers to ``\textit{design or implementation constructs that are expedient in the short term, but set up a technical context that make future changes more costly}.'' On the implementation level, such constructs are referred to as code smells. There are several tools available to detect code smells, such as CAST, SonarQube, and CodeScene. Developers typically ``pay back'' TD through refactoring after they detect code smells. Unfortunately, market pressure typically leaves little room for refactoring, causing code quality degradation and systemic risks. This directly undermines two types of sustainability as defined in the ``Karlskrona Manifesto for sustainability design''~\cite{becker_karlskrona_2015}, namely: \textit{technical sustainability} (by hindering code evolution as new features are needed) and \textit{economic sustainability} (by threatening long-term competitiveness).

Beyond its technical and economic impact, TD also has an important human aspect. Work by Graziotin and colleagues identified a strong connection between code quality and developer happiness. In a large-scale quantitative and qualitative survey, they found that bad code quality and coding practice are leading causes of developer unhappiness~\cite{graziotin_unhappiness_2017}. Furthermore, they report that developer unhappiness leads to low productivity, low code quality, low motivation, and work withdrawal~\cite{graziotin_what_2018}. In other words, unmaintainable code increases the risk that developers quit, which ties into a third sustainability type in the Karlskona Manifesto: \textit{individual sustainability} -- retaining human capital in an organization. One could even argue that TD raises ethical concerns about developer well-being and retention.

Despite the importance of maintainability, many studies report that refactoring tasks rarely get the development budget they deserve. One major reason, reported in a systematic literature review by Li \textit{et al.}~\cite{li_systematic_2015}, is a gap between business and engineering perspectives. It is hard for organizations to assign a business value to maintainability, as it is an internal product quality. But given its multi-faceted importance for sustainability, omitting the activity due to this challenge is an irresponsible decision. Instead, this highlights a need for additional decision-support models to help CTOs, tech leads, product managers and other roles to set appropriate code-level quality targets for maintainability. Taking responsibility for maintainability has the potential to support multiple direct and indirect stakeholders, including 1) customers who expect future software evolution, 2) shareholders who invest in long-term market success, and 3) internal developers who need it for work satisfaction. 

We argue that modern tool support and Requirements Engineering (RE) could help organizations responsibly set appropriate maintainability targets. As discussed in Section~\ref{sec:rw}, RE research on maintainability has been very limited. In previous work, we speculated about the potential value of quantitative maintainability requirements in various scenarios, including outsourcing, open-source component selection, company acquisitions, and public procurement~\cite{borg_quality_2024}. Furthermore, we hypothesized that the QUPER model, developed by Regnell \textit{et al.}~\cite{regnell_supporting_2008} for performance requirements, could be adapted for maintainability requirements. We now present a first solution proposal based on an ongoing design science process.

This paper summarizes a MSc thesis~\cite{breid_improving_2025} that makes two contributions. First, we explore how maintainability requirements are addressed in industry practice. Second, we propose QUPER-MAn as an adaptation of QUPER tailored for maintainability. We have completed one full build-evaluate loop followed by a second build activity. An important component of QUPER-MAn is the inclusion of benchmarking data, as maintainability among industry competitors is not an observable quality -- in contrast to performance as addressed by QUPER.

The rest of this paper is structured as follows. Section~\ref{sec:rw} presents background and related work. Section~\ref{sec:method} presents our study design consisting of two sequential phases. In Section~\ref{sec:res}, we share our results and discuss the findings. Finally, we conclude the paper in Section~\ref{sec:conc} and outline the most important directions for future work. 

\section{Background and Related Work} \label{sec:rw}
This section first introduces CodeScene, CodeHealth and our previous work on maintainability. Second, we present related work on quality requirements in general and QUPER in particular.

\subsection{CodeScene, CodeHealth, and Maintainability}
CodeScene is a quality and TD management tool that uses static code analysis with behavioral data from version control systems to identify maintainability issues~\cite{tornhill_prioritize_2018}. The tool identifies and highlights TD areas that might hinder developers' comprehension and maintenance of software projects across 31+ programming languages. CodeScene uses a maintainability metric, CodeHealth, to quantify and visualize such issues, which helps organizations understand the state of their projects and where to prioritize refactoring efforts.

Numerous papers have proposed code quality metrics for maintainability~\cite{riaz_systematic_2009,baggen_standardized_2012}. Such metrics are commonly used to estimate how difficult it is for a developer to make changes to the corresponding piece of code. A naïve metric that often works surprisingly well on both the file and method level is to simply count the Lines of Code (LoC) -- bigger volumes of code are generally seen as harder to maintain by humans. Some metrics focus on complexity, such as Cyclomatic Complexity and Halstead Complexity metrics, or object-oriented concerns such as depth of inheritance trees and class coupling. An early aggregate metric, the maintainability index~\cite{oman_metrics_1992}, combines LoC, Cyclomatic and Halstead Complexity, and the percentage of code comments. Additionally, there are tool-specific metrics based on code smell detectors, e.g., CodeScene's CodeHealth™ and SonarQube's TD Index.

CodeHealth aggregates 25+ code smells known to challenge humans' program comprehension into a score from 1 to 10 where high scores indicate maintainable code. We have validated CodeHealth's association with defect density and implementation time using a large dataset  of 39 proprietary projects~\cite{tornhill_code_2022}. Two years later, we extended the dataset to roughly twice its size and conducted a more detailed analysis of these associations~\cite{borg_increasing_2024}. In that study, we also presented a corresponding value model that acts as a cornerstone for this work. Furthermore, we have compared CodeHealth, SonarQube's TD Index, and maintainability index in a benchmarking study~\cite{borg_ghost_2024} on the human-annotated maintainability dataset provided by Schnappinger \textit{et al.}~\cite{schnappinger_defining_2020}. As CodeHealth outperformed the alternatives, we rely on this metric in this study.

\subsection{Quality Requirements and QUPER}
Numerous studies highlight Quality Requirements (QR) as particularly challenging. Berntsson Svensson \textit{et al.} report that QRs are often poorly understood, stated informally without quantifiable targets, and thus cannot be validated~\cite{berntsson_svensson_quality_2011}. Furthermore, based on a study with 11 companies, they found that about 20\% of all QRs are dismissed during development -- and internal qualities such as maintainability are more often dismissed. In another study, based on an analysis of QRs sent to suppliers, Berntsson Svensson \textit{et al.} state that maintainability has few quantified QRs~\cite{berntsson_svensson_investigation_2013}. Therefore, they elaborate, the QUPER model (described next) is inadequate unless ``\textit{a structured and systematic method may help in achieving the quantification.}'' We argue that CodeHealth offers the missing piece for code-level maintainability requirements.

The QUPER model, short for quality performance, was created by Regnell \textit{et al.} to support QR management in an increasingly market-driven software engineering context~\cite{regnell_supporting_2008}. It was developed in several stages in close collaboration with industry partners, following the Technology Transfer Model (TTM) by Gorschek \emph{et al.}~\cite{gorschek_model_2006}. As recommended by TTM, QUPER has been both statically and dynamically validated several times~\cite{ berntsson_svensson_case_2015}. The QUPER version described in this paper is the model that evolved as a result of these validation studies. 

The QUPER creators report that the model simplifies the prioritization and roadmapping of QRs by making tradeoffs between cost, benefit, and quality explicit. It was originally developed for the smartphone domain, for which it has been complemented with practical guidelines~\cite{berntsson_svensson_case_2015}. For performance requirements, QUPER has proven to be useful also for the development of financial services~\cite{berntsson_svensson_setting_2012}. Moreover, we have successfully used QUPER to discuss the benefit of different quality levels in evaluative studies related to the accuracy of recommendation systems~\cite{borg_adopting_2024}, indicating that the model can be generalized to various contexts. Our positive experience motivates us to pursue an adapted version of QUPER for yet another quality aspect: maintainability.

QUPER provides three complementary views that support high-level decision-making related to quality: 1) benefit, 2) cost, and 3) roadmap. At the core are the concepts of \textit{breakpoints} and \textit{barriers} along with the assumption that quality levels have non-linear relationships with both benefit and cost.

\begin{itemize}
    \item The \textbf{Benefit view} shows the relation between quality and benefit in terms of three breakpoints. Below the \textit{Utility} breakpoint, users do not recognize the value of the solution. The \textit{Differentiation} breakpoint marks the shift from useful to competitive quality, i.e., users get a ``wow’’ feeling. Finally, quality levels beyond the \textit{Saturation} breakpoint do not provide any additional value. The dashed line in Figure~\ref{fig:benefit_v0_9} depicts the relationship.
    \item The \textbf{Cost view} (see Figure~\ref{fig:cost_v0_9}) illustrates the relation between quality and cost using barriers. Barriers appear when an increase in quality has a substantial cost penalty. 
    \item The \textbf{Roadmap view} combines the elements from the benefit and cost views into a single axis. This view also allows for the inclusion of competitors' solutions for comparison.    
\end{itemize}

Even though there is extensive research on maintainability software metrics, RE research on maintainability is very limited~\cite{borg_quality_2024}. In response to our previous call for research~\cite{borg_requirements_2023}, we now present QUPER-MAn -- a preliminary solution proposal combining QUPER and CodeHealth.

\section{Method} \label{sec:method}
The long-term goal of this research is to help organizations responsibly manage software maintainability. The current work is guided by two research questions:

\begin{itemize}
    \item[RQ1] How are maintainability requirements currently handled in industry practice?
    \item[RQ2] How can the QUPER model be adapted to support target setting for source code maintainability requirements?
\end{itemize}

\begin{figure}
  \centering
  \includegraphics[width=1\linewidth]{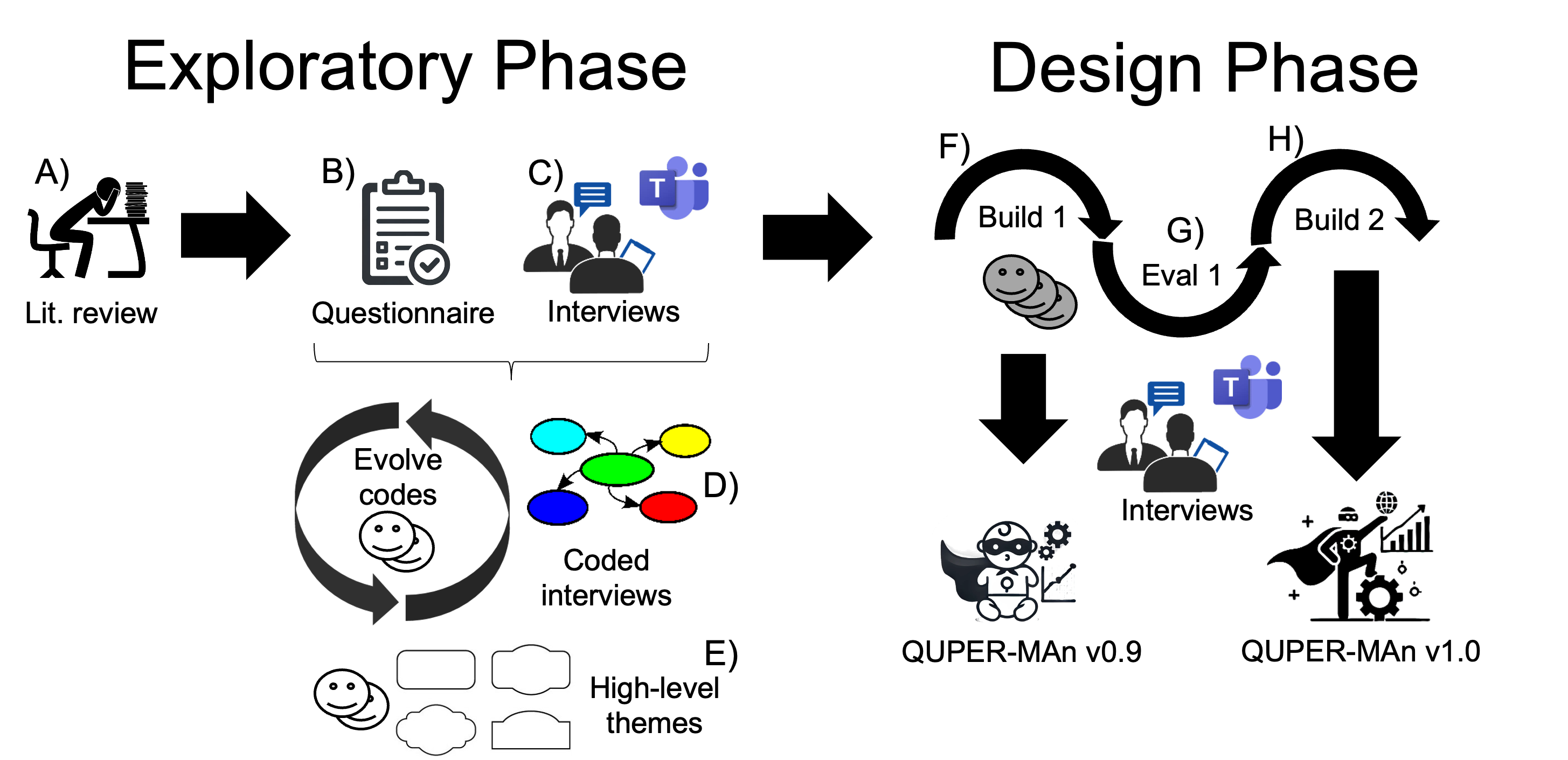}
  \caption{Method overview. White smileys indicate authors of this paper, gray shows external experts.}
  \label{fig:method}
\end{figure}

As presented in Figure~\ref{fig:method}, the study design is two-fold. First, we conducted an exploratory study to address RQ1, involving a questionnaire-based survey and in-depth interviews. Second, related to RQ2, we commenced engineering research which we frame as design science with alternating \textit{build} and \textit{evaluate} steps~\cite{hevner_design_2004}. This paper presents the results from a complete design cycle followed by an improving design step. A detailed description of the method is available in Breid and Larsson~\cite{breid_improving_2025}.

\subsection{Exploratory Phase}
\textbf{A) Literature review.} We initiated the work by reviewing academic papers as summarized in Section~\ref{sec:rw}.

\noindent \textbf{B) Questionnaire-based survey}. We designed a questionnaire using the SUNET survey engine to explore the state of practice. To increase the chances of a good response rate, we opted for a very short survey (<10 min) as can be seen in the Appendix~\ref{app:quest} with five demographic questions (D1-D5), two Likert scales (L1-L2), one closed question about maintainability (C1), and one open question (O1) about how maintainability specifications are currently specified. Finally, we asked if the respondent was open to a follow-up interview. Five practitioners volunteered to validate the survey, which led to slight refinements.

First, we invited CodeScene customers to answer the survey. Despite the short questionnaire and targeting a population who actively pays for maintainability support, the response rate was very low. We resorted to also share invitations on CodeScene's social media and our personal contact networks. Still, the total number of respondents only reached 29. Despite the low response rate, we analyzed the collected data using descriptive statistics. Additionally, two authors jointly conducted thematic coding as will be described in the next section.

\noindent \textbf{C) Interviews}. We created an interview guide to collect rich information about maintainability requirements in a semi-structured format. The goal was to understand current practices, tools, and metrics used to support maintainability and manage TD. Our particular focus was on how they elicit, specify, prioritize, and validate maintainability requirements -- if that type of QR was at all considered in the development process. We also explored current challenges in relation to maintainability. To validate the interview guide, we conducted a pilot interview with a CodeScene employee.

The interviews (N=5) were conducted using MS Teams and we used its internal transcription service followed by manual corrections and member checking. We did a thematic analysis of the interview data adhering to the five-step procedure prescribed by Cruzes and Dybå (2011) \cite{cruzes_recommended_2011}. 1) We familiarized ourselves with the data during the cleaning process. 2) Two authors independently coded the transcripts -- and the free-text answers from the survey -- and iteratively evolved a coding guide. 3) Codes were iteratively grouped into themes. 4) The themes were organized into three groups, again in an iterative manner. 5) We assessed the validity of the synthesis. Due to this paper's strict page limit, we report only key takeaways -- details are found in~\cite{breid_improving_2025}.

\subsection{Design Phase}
\begin{figure*}[ht]
    \centering
    \begin{subfigure}[t]{0.3\textwidth}
        \includegraphics[width=\textwidth]{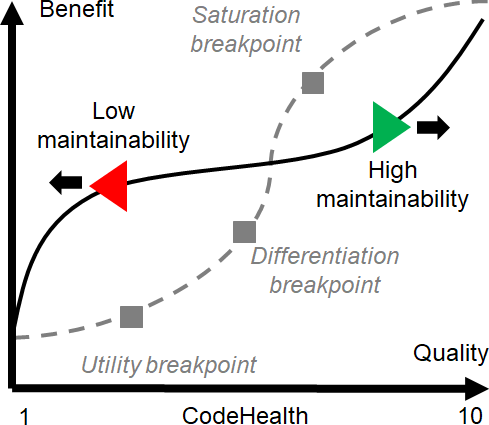}
        \caption{Benefit view.}
        \label{fig:benefit_v0_9}
    \end{subfigure}
    \hspace{0.03\textwidth} 
    \begin{subfigure}[t]{0.3\textwidth}
        \includegraphics[width=\textwidth]{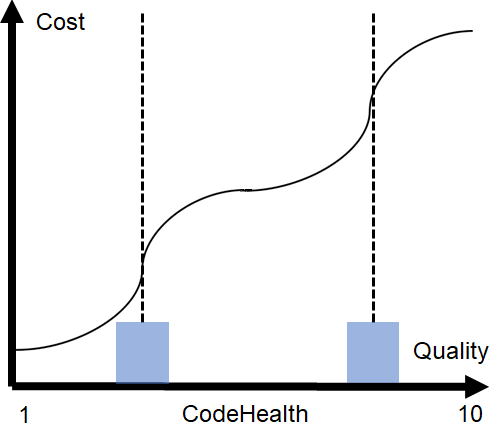}
        \caption{Cost view.}
        \label{fig:cost_v0_9}
    \end{subfigure}
    \hspace{0.03\textwidth} 
    \begin{subfigure}[t]{0.3\textwidth}
        \includegraphics[width=\textwidth]{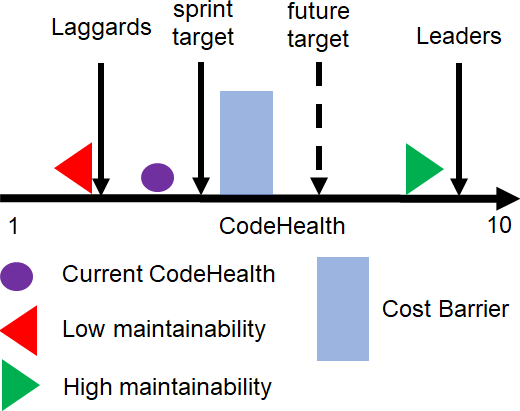}
        \caption{Roadmap view.}
        \label{fig:roadmap_v0_9}
    \end{subfigure}
    \caption{QUPER-MAn v0.9. The dashed line in the Benefit view shows the shape of the original QUPER model and its three breakpoints. The Cost view, showing two cost barriers, is identical to the original QUPER counterpart. The Roadmap view shows example data for a project with low CodeHealth. The dashed vertical arrow depicts a future CodeHealth target that requires passing a cost barrier.}
    \label{fig:quperman_v0_9}
\end{figure*}

\textbf{F) Build iteration 1}. In the first build iteration, we adapted QUPER to better match maintainability specifics. As part of the iteration, we invited three customer-facing CodeScene employees to provide early feedback on the design. QUPER-MAn v0.9 introduced CodeHealth on the X-axis as a quantitative maintainability score as well as the following changes to QUPER:

\begin{itemize}
    \item Benefit view: As shown in Fig.~\ref{fig:benefit_v0_9}, we replaced QUPER's sigmoid curve shape with the shape of the value curve from our previous work~\cite{borg_increasing_2024} resembling a logit function. Moreover, we replaced the three QUPER breakpoints with two new breakpoints from which changes are non-linear.
    \item Cost view: We kept this view intact but created an initial list of possible cost barriers for maintainability improvements, including architectural changes, test case development, and revision of documentation. We intended to elicit additional barriers during the evaluation step.
    \item Roadmap view: Since competitors' maintainability levels are not directly observable, in contrast to certain aspects of performance, we replaced QUPER’s vertical competitor arrows with benchmarking data. The data originates from data from CodeScene’s customer pool~\cite{borg_industrial_2025}, and can be filtered to match particular market niches, programming languages, company sizes, etc. Using this data, Fig.~\ref{fig:cost_v0_9} positions \textit{Leaders} and \textit{Laggards} on the X-axis, representing the 90th and 10th percentiles, respectively.
\end{itemize}

\textbf{G) Evaluation 1}. We conducted three evaluative interviews with industry practitioners to discuss our design proposal. One of the interviewees from the exploratory phase was again interviewed, the other two were new contacts. The interviews were conducted with MS Teams and consisted of three parts: 1) An introduction to the original QUPER model, 2) An introduction to CodeScene and CodeHealth, including the benchmarking data, and 3) A presentation of QUPER-Man v0.9. The feedback elicitation was conducted under a flexible think-aloud protocol, and we particularly focused on eliciting input for the Cost view. We recorded the interviews and used an automatic transcription service. 

\textbf{H) Build iteration 2}. Based on the feedback in the evaluation step, we created QUPER-MAn v1.0. The major changes to the solution proposal, described in detail in Section~\ref{sec:rq2}, were as follows:
\begin{itemize}
    \item Benefit view: We renamed the two breakpoints to the \textit{Cost Spiral Trigger} and the \textit{Value Cascade Point}.
    \item Cost view: We replaced the solution with cost barriers positioned at specific CodeHealth values with investment thresholds at a relative distance from the current CodeHealth level. Moreover, we added expected future thresholds at equidistant future levels. This is further justified in Section~\ref{sec:rq2}.
    \item Roadmap view: We reflected the changes to the other views in the Roadmap view. Furthermore, as presented in Fig.~\ref{fig:quperman_v1_0}, we added a quality gate and an ``escalation zone'' denoting the increasing cost of CodeHealth improvements until an investment barrier is reached. 
\end{itemize}

\section{Results and Discussion} \label{sec:res}
This section discusses our findings related to the two RQs.

\subsection{RQ1: Maintainability Requirements in Practice}
Despite high ambitions, we only collected 29 responses to our survey. Due to the low number of responses, the data has not been made publicly available but can be shared upon request. We acknowledge the limited generalizability of the questionnaire-based findings of the research. Adhering to a flexible study design, we instead largely resorted to in-depth interviews.

\begin{figure}
  \centering
  \includegraphics[width=1\linewidth]{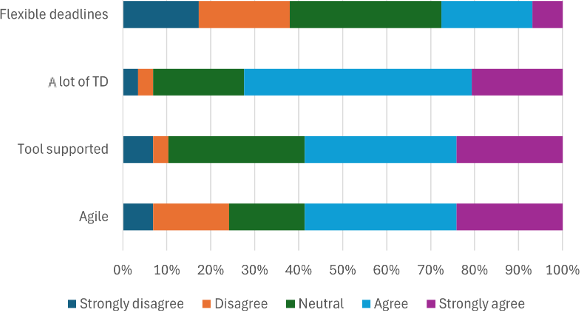}
  \caption{Development characteristics.}
  \label{fig:context}
\end{figure}

Despite the low survey response rate, we provide a brief overview of the respondent demographics. Most of our respondents self-reported working in the Computer (Software) domain (D1) with product development (D3) in large organizations (D5). All respondents had at least 5 years of professional experience, and the median answer was 10-19 years (D2). Figure~\ref{fig:context} shows the Likert scale related to statements about the development context (L1). We find that 1) projects are differently strict regarding deadlines, 2) high levels of TD are reported by 21 respondents (72\%), 3) 17 respondents report working with modern tools (59\%), and 4) agile contexts are roughly twice as common as plan-driven development.

Five practitioners from different organizations agreed to follow-up interviews to elaborate on the topic. As presented below, the five interviewees represent a diverse set of development contexts.
    
\begin{itemize}
    \item[\textbf{Int1}] A lead developer in a large company primarily focusing on software for the public sector. The interviewee develops and leads features while championing maintainability within the projects.
    \item[\textbf{Int2}] A lead developer and ``technical evangelist'' at a small company providing full-service solutions for the financial industry. The company specializes in portfolio management and provides middle and back-office solutions for investment companies and investment funds. 
    \item[\textbf{Int3}] A management consultant who works for a multinational retail organization as an Enterprise Resource Planning (ERP) service manager. ERP systems are integrated IT platforms that centralize and automate core business processes, such as finance, supply chains, and operations.
    \item[\textbf{Int4}] A product owner and requirements engineer working for a large multinational company. The interviewee's team focuses on developing software for various displays as components for electric vehicles.
    \item[\textbf{Int5}] A self-employed solo developer delivering bespoke software for 10+ customers.
\end{itemize}

\begin{figure}
  \centering
  \includegraphics[width=1\linewidth]{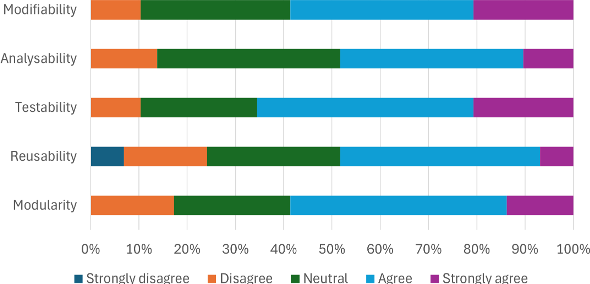}
  \caption{Importance of the maintainability sub-categories.}
  \label{fig:subcats}
\end{figure}

Figure~\ref{fig:subcats} shows to what extent the respondents indicated that the different ISO/IEC~25010's maintainability sub-categories were important in their organizations (cf. Appendix L2). We notice that the respondents agreed that all sub-categories were important, i.e., the differences between them are small. 

\begin{figure}
  \centering
  \includegraphics[width=0.85\linewidth]{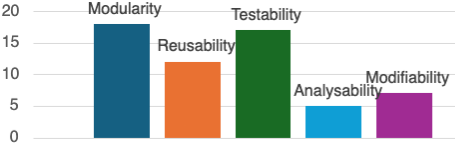}
  \caption{Most challenging maintainability sub-categories.}
  \label{fig:bars}
\end{figure}

Figure~\ref{fig:bars} indicates that \textit{modularity} and \textit{testability} were considered the most challenging by the survey respondents. Int2 presented an example that combines the two aspects, explaining how challenging it is to migrate a monolithic (non-modular) legacy system with poor testability to a microservice architecture. 

Several free-text responses from the survey suggest that many organizations consider maintainability important, but not a top priority -- in line with previous work~\cite{berntsson_svensson_quality_2011}. One survey response captures the sentiment of many respondents and interviewees: \emph{``The management doesn't see value in considering or planning for maintainability, believing developers should just push out new code as fast as possible. The business analysts and product owners aren't technical and have no firm understanding of maintainability /\ldots/ that never appears in requirements. It's just something developers are somehow supposed to make happen in a hurry to drop new code''.}

Most interviewees were satisfied users of static code analysis tools for code smell detection. However, as commonly reported in the TD literature~\cite{li_systematic_2015,ernst_technical_2021}, time pressure and short-term goals often lead to deprioritization of acting on the identified smells. \textbf{Int5} proves that this can be the case also for a solo developer: \emph{My biggest enemy is time. And it's the old paradox /\ldots\/ You're too busy sawing down trees to stop and sharpen your sword. Because if you stop and sharpen your sword, you're not sawing down trees. /\ldots/ I'd love to write beautifully formatted, maintainable code. But the problem is everybody wants something in a hurry so they don't want to wait two days or three days or a week for something.}

Apart from tool-driven quality gates from linters and code smell detectors -- ``don’t make it worse’’ -- none of the respondents or interviewees shared any explicit approaches to eliciting quantitative targets for maintainability requirements. Instead, maintainability requirements commonly refer to vague coding conventions that developers should follow. Beyond quality gating, neither respondents nor interviewees mentioned any particular maintainability-focused RE processes. Instead, it is typically up to tech leads, architects, and other senior roles to prioritize refactoring efforts ad-hoc and to request budgets accordingly. \textbf{Int4} explained it as:  \emph{``if the architect tells us to clean up some part of the code, then we do so. But that's not connected to the requirements.''}

Based on findings from the exploratory phase, including the literature review, we identified two key goals for the design phase. First, maintainability requirements must become a first-class citizen for responsible software engineering, with discussions involving both technical and business stakeholders. Second, maintainability requirements need to be quantified to ensure they are actionable and verifiable. In the design phase, we relied on QUPER to address the first goal and CodeHealth for the second. For a longer discussion, we refer to Breid and Larsson~\cite{breid_improving_2025}.

\subsection{RQ2: Solution Proposal} \label{sec:rq2}
\begin{figure*}[ht]
    \centering
    \begin{subfigure}[t]{0.27\textwidth}
        \includegraphics[width=\textwidth]{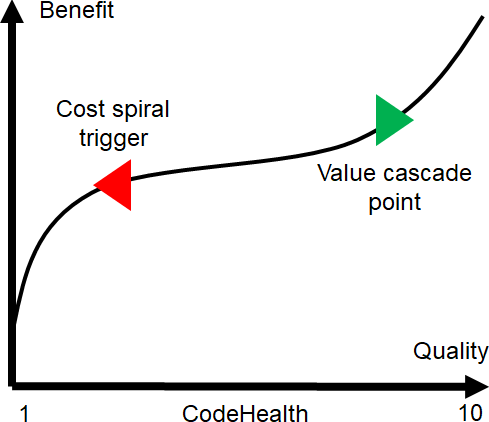}
        \caption{Benefit view.}
        \label{fig:benefit_v1_0}
    \end{subfigure}
    \hspace{0.03\textwidth} 
    \begin{subfigure}[t]{0.29\textwidth}
        \includegraphics[width=\textwidth]{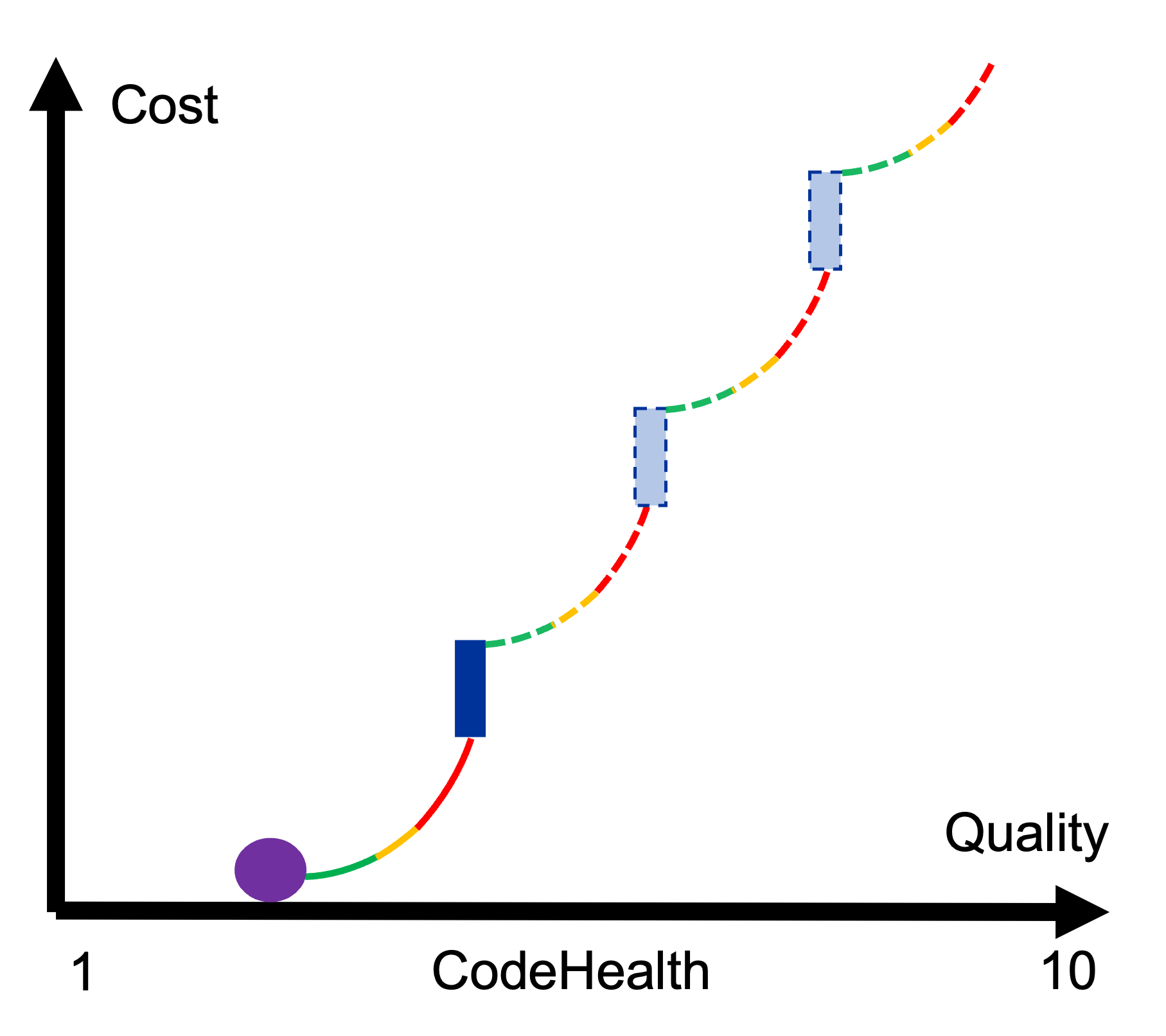}
        \caption{Cost view.}
        \label{fig:cost_v1_0}
    \end{subfigure}
    \hspace{0.03\textwidth} 
    \begin{subfigure}[t]{0.35\textwidth}
        \includegraphics[width=\textwidth]{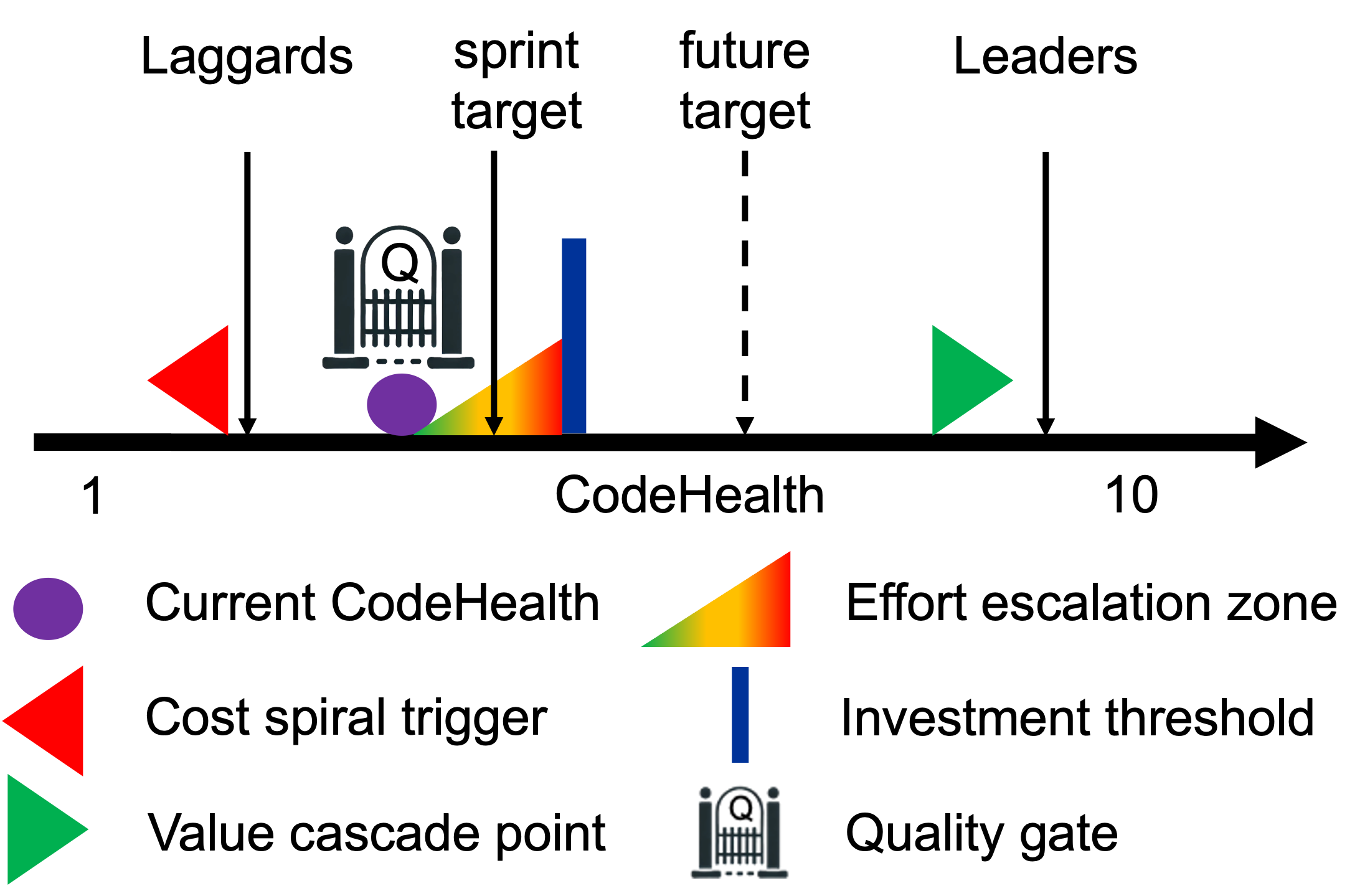}
        \caption{Roadmap view.}
        \label{fig:roadmap_v1_0}
    \end{subfigure}
    \caption{QUPER-MAn v1.0. The Benefit view shows the value of maintainability. The Cost view shows the escalating costs of refactoring for maintainability with investment barriers that need to be overcome. The Roadmap view has been extended with an effort escalation zone, and investment barrier, and a quality gate.}
    \label{fig:quperman_v1_0}
\end{figure*}

Figure~\ref{fig:quperman_v1_0} shows the three views of QUPER-MAn v1.0. All three views have CodeHealth on the X-axis. 

The \textit{Benefit view} shows a curve characterized by the \textit{Cost Spiral Trigger} and the \textit{Value Cascade Point}. Between the points, the relationship between quality and benefit is linear. Towards the extremes of the CodeHealth range, non-linearities manifest. The shape of the curve is the inverse of the counterpart in the original QUPER model, based on empirical findings on the value of different CodeHealth scores~\cite{borg_increasing_2024}.

The circle in the \textit{Cost view} indicates the current CodeHealth level. First, the costs of improving CodeHealth is low, but as the low-hanging fruits have been collected, the refactoring costs turn gradually more expensive (cf. the green--yellow--red curve). At some point, there will be an investment threshold that needs to be overcome -- which will be costly and perhaps not directly influence the CodeHealth. Once such a major refactoring endeavor has been implemented, it is expected that a new starting point has been reached and the pattern repeats as highlighted using dashed curves and thresholds. During our interviews, we elicited a list of potential investment thresholds by asking interviewees to brainstorm and ``think aloud.'' As part of Build iteration 2, we refined these into the following items:

\begin{description}
    \item[Test adequacy] Refactoring without fear of breaking functionality requires reliable test coverage. Organizations might need to invest in improved test suites and test infrastructure before refactoring can take place. Ensuring adequate tests before making changes is an established reengineering best practice~\cite{demeyer_object-oriented_2002}.
    \item[Architectural change] Some refactorings necessitate changes to the system architecture, which are inherently costly. As Grady Booch famously expresses it: ``Architecture represents the significant design decisions that shape the form and function of a system, where significant is measured by the cost of change.''\footnote{\url{https://handbookofsoftwarearchitecture.com/}} Examples of major architectural adjustments include microservice migration and the adoption of an event-driven architecture.  
    \item[Developer Training] Certain refactorings might require replacement of outdated technology through adoption of modern tools and techniques. Developers might need upskilling efforts and training to handle changes to the technology stack. Examples include new frameworks such as Spring Boot or React, container orchestration with Kubernetes, and Cassandra for distributed databases.
    \item[Knowledge recovery] Critical system knowledge, particularly in legacy systems, may have been lost over time due to staff turnover or inadequate documentation. Recovering this knowledge can require reverse engineering to rediscover how the system functions.
    \item[Domain knowledge] Refactoring certain parts of a system may require specialized domain knowledge that no longer exists within the organization. Addressing these gaps might involve conducting research or consulting external experts to gain the necessary understanding.
    \item[Regulatory aspects] Code that must comply with legal or regulatory requirements adds an extra layer of complexity. Ensuring compliance after refactoring may involve major test campaigns, legal reviews, or even re-certification processes. The latter is acknowledged as a major obstacle when evolving safety-critical systems~\cite{de_la_vara_amass_2019}.
    \item[Major code smells] \textit{God Classes} and \textit{Brain Methods} can be particularly challenging to modularize. Both of these smells take on too many responsibilities, making the code difficult to understand, modify, and modularize. Untangling such code smells, especially when intertwined with dense dependencies, can significantly slow progress. Modularization efforts may require extensive planning and careful execution.
\end{description}

\subsection{Threats to Validity}
The work presented in this study is part of a long-term effort to develop decision support for maintainability target-setting.  Below, we outline the most pressing threats to the validity of our conclusions and the feasibility of QUPER-MAn v1.0.

\textbf{Construct validity.} The survey and the interviews focus on the non-trivial maintainability construct. We relied on the ISO/IEC~25010 definition of maintainability, which we argue is both \textit{valid} and sufficiently \textit{complete}, given its five sub-categories. The main threat is  whether study participants interpreted the construct consistently. To mitigate this, we provided explanatory text in the questionnaire and carefully structured introductions during the interview sessions.

\textbf{Reliability.} Would other researchers reach similar conclusions given the same data? Interpretation is inherent in qualitative research, and the two researchers responsible for the thematic analysis have limited engineering experience. To mitigate potential bias, we followed standard practice including joint development of research instruments, pilot testing, member checking, and thematic analysis. These measures increased alignment between researchers. Importantly,  there was no single-coder bias.

\textbf{Internal validity.} QUPER-MAn insinuates causal relationships, most prominently about CodeHealth and Benefit. While this association is supported by prior regression analysis~\cite{borg_increasing_2024}, software engineering is a complex activity with numerous confounding factors at play. Moreover, we suggest that working with QUPER-MAn could help organizations improve their maintainability management. However, this must be evaluated through longitudinal case studies in real-world industrial settings. 

\textbf{External validity.} We initially aimed for a large survey of maintainability requirements in industry practice, but our sampling strategy failed. Instead, we shifted to collecting rich data from interviews. While we cannot make strong claims about the absence of maintainability requirements based on our work, our findings echo previous work -- maintainability is often a consequence rather than an active decision. The three interviewees discussing QUPER-MAn represent different development contexts, but the sample size is too small for generalization. Future work will involve additional build-evaluate cycles to further assess feasibility.

\section{Conclusion} \label{sec:conc}
Our exploratory study corroborates previous work, showing that many organizations perceive maintainability as a nice-to-have rather than a must-have quality. Furthermore, in line with the TD literature, study participants highlighted the perceived value of short time-to-market over long-term maintainability. The irresponsible focus on short-term gains at the expense of long-term value -- known as \textit{hyperbolic discounting} in behavioral economics -- is certainly present in software engineering. This is a costly problem, and various approaches are needed to inform the inevitable trade-off. 

Tool support for maintainability is prevalent in industry practice. However, our study reveals a limitation in how organizations set targets based on tool output quality levels. Current practices appear to be largely restricted to quality gating and ad hoc improvement campaigns. This indicates a potential for structured requirements engineering to set appropriate targets in a responsible manner. Moreover, as tools are widely adopted, they form a natural component of the solution. In our solution proposal, we advocate using CodeHealth for quantified targets, as it outperforms alternative metrics on the most reliable benchmark with human maintainability assessments~\cite{borg_ghost_2024}.

This preliminary study culminates in the first version of QUPER-MAn, an adaptation of the established QUPER model. Key adaptations to fit the maintainability context include: 1) the \textit{Benefit view} features an inverted curve with two breakpoints, 2) the \textit{Cost view} introduces investment thresholds at a relative distance from the current quality level, and 3) the \textit{Roadmap view} introduces benchmark values for industry ``leaders'' and ``laggards'' to enable comparisons.

As this is a work in progress, future work will follow the same design science trajectory with additional build-evaluate cycles. We plan to conduct future interviews with industry practitioners to validate the QUPER-MAn feasibility. After subsequent refinements, we envision industrial case studies to evaluate how the model stimulates discussions regarding appropriate maintainability levels for software projects at different lifecycle stages. Furthermore, we consider creating a version that is more inspired by the TD concepts of principal and interest.

\section*{Acknowledgement}
This work was partially supported by the Competence Centre NextG2Com funded by the VINNOVA program for Advanced Digitalisation with grant number 2023-00541 and partly by the Wallenberg AI, Autonomous Systems and Software Program (WASP).